\documentstyle[12pt,epsfig]{article}
\begin{document}
\def\scri{\unitlength=1.00mm
\thinlines
\begin{picture}(3.5,2.5)(3,3.8)
\put(4.9,5.12){\makebox(0,0)[cc]{$\cal J$}}
\bezier{20}(6.27,5.87)(3.93,4.60)(4.23,5.73)
\end{picture}}

\begin{center}

{\Large ANTI-DE SITTER QUOTIENTS,}

\vspace{8mm}

{\Large  BUBBLES OF NOTHING, AND BLACK HOLES}

\vspace{15mm}

{\large Jan E. \AA man}$^*$\footnote{Email address: ja@physto.se}

\

{\large Stefan \AA minneborg}$^{**}$\footnote{Email address: 
stefan.aminneborg@utbildning.stockholm.se}

\

{\large Ingemar Bengtsson}$^*$\footnote{Email address: ingemar@physto.se. 
Supported by VR.}

\

{\large Narit Pidokrajt}$^*$\footnote{Email address: narit@physto.se}

\vspace{8mm}

* {\sl Stockholm University, AlbaNova\\
Fysikum\\
S-106 91 Stockholm, Sweden}

\vspace{5mm}

** {\sl Norra Reals Gymnasium\\
S-113 55 Stockholm, Sweden}

\vspace{10mm}

{\bf Abstract}

\end{center}

\vspace{5mm}

\noindent In 3+1 dimensions there are anti-de Sitter quotients which 
are black holes with toroidal event horizons. By analytic continuation of 
the Schwarzschild-anti-de Sitter solution (and appropriate identifications) 
one finds two one parameter families of spacetimes that contain these quotient 
black holes. One of these families consists of B-metrics (``bubbles of 
nothing''), the other of black hole spacetimes. All of them have vanishing 
conserved charges.

\vspace{5mm}

\newpage

{\bf 1. Introduction}

\vspace{5mm} 

\noindent Because of the special properties of the conformal boundary of 
anti-de Sitter space, one can form black hole spacetimes by taking the 
quotient of anti-de Sitter space with discrete isometry groups. 
In 3+1 dimensions there are two classes of such black holes: static 
``topological'' black holes, and a non-stationary example with a toroidal 
event horizon \cite{4d}. The former acquire much of their interest as 
members of a one parameter family of asymptotically anti-de Sitter black 
holes \cite{BLP}. The latter can be regarded as a special case of the   
Ehlers-Kundt B1 metrics \cite{Ehlers}---also known as ``Kinnersley 
type IV.B'' \cite{Kin}, or more catchily as  ``bubbles of nothing'' 
\cite{Witten,Aharony}---but these spacetimes do not describe black holes 
except in one very special case. In fact they are everywhere regular 
spacetimes. The main purpose of this paper is to 
point out that there is another one parameter family of spacetimes 
that does describe black holes for all values of the parameter, and 
which also contains the toroidal quotient black hole. Although they are 
free of curvature singularities, they are not everywhere regular. 
All three families can be locally obtained through analytic 
continuation from the Schwarzschild-anti-de Sitter spacetimes. The two 
families that we will discuss share the property that all their 
conserved charges, in the 
sense of Ashtekar and Magnon \cite{Ashtekar}, are zero. 

In section 2 we will recapitulate the properties of the toroidal anti-de 
Sitter black hole. In section 3 we present spacetimes analytically 
related to Schwarzschild. In section 4 we discuss the bubbles of nothing, 
and in section 5 the black hole family. Section 6 provides a summary. 

\vspace{1cm}

{\bf 2. A quotient black hole}

\vspace{5mm}

\noindent Anti-de Sitter space (or adS) is a spacetime with constant 
non-zero curvature, conveniently regarded as the covering space of the 
quadric surface 

\begin{equation} X^2 + Y^2 + Z^2 - U^2 - V^2 = - 1 \end{equation}

\noindent in a flat space with the metric 

\begin{equation} ds^2 = dX^2 + dY^2 + dZ^2 - dU^2 - dV^2 \ . \end{equation} 

\noindent The intrinsic metric can be given as  
 
\begin{equation} ds^2 = - \left( \frac{1 + \rho^2}{1-\rho^2}\right)^2 dt^2 + 
\frac{4}{(1-\rho^2 )^2}\left( d\rho^2 + \rho^2(d\theta^2 + \sin^2{\theta}d\phi^2)
\right) \ , \label{salami} \end{equation}

\noindent where 

\begin{equation} 0 \leq \rho < 1 \ , \hspace{8mm} 0 < \theta < \pi \ , 
\hspace{8mm} 0 \leq \phi < 2\pi \ . \end{equation}

\noindent Anti-de Sitter space is conformal to ``one half'' of the Einstein 
universe, and one can add a conformal boundary at $\rho = 1$. 
%
%, if one works with the conformally related metric 
%
%\begin{equation} d\hat{s}^2 = - dt^2 + \frac{4}{(1+\rho^2)^2}
%\left( d\rho^2 + \rho^2(d\theta^2 + \sin^2{\theta}d\phi^2)\right) \ . 
%\label{ghatt} \end{equation}
%   
The conformal boundary is denoted \scri , and is itself a conformal copy of 
the Einstein universe in one dimension less. 

In 3+1 dimensions all possible spacetimes arising by performing identifications 
using one-parameter subgroups of $SO(3,2)$ have been classified \cite{holstpeldan, 
I}. A black hole is obtained when the subgroup is generated by the Killing vector 

\begin{equation} \xi = J_{ZU} = Z\partial_U + U\partial_Z \ . \end{equation}

\noindent At fixed $t$ the resulting spacetime can be described as the 2+1 
dimensional spinless BTZ black hole \cite{BHTZ}, rotated around a spatial 
axis. 

The details are explained elsewhere \cite{4d, holstpeldan, Mann, I}, but 
since it is an interesting black hole we will pause to understand it. 
Its global isometries are 
generated by anti-de Sitter Killing vectors commuting with $J_{ZU}$. They 
form the Lie algebra $SO(2)\times SO(2,1)$. A Misner 
singularity terminates \scri , which consists of a single component. In 
covering space, the event horizon is at 

\begin{equation} X^2 + Y^2 - V^2 = 0 \ . \end{equation}

\noindent Points with $X^2 + Y^2 - V^2 < 0$ are invisible from \scri , and will 
therefore lie in the interior of a black hole. There is a Misner singularity 
also in the physical spacetime---in covering space it is a 
disk at $Z = U = 0$, which is behind the horizon. The horizon itself has some 
interesting properties. 
At fixed time $t > 0$ it is a torus. One of its cycles is growing, 
but there is another cycle of constant length. 
To see why this is so, recall that a Killing horizon is a light cone with a 
vertex on \scri ---its generators do not diverge because its vertex is 
infinitely far away. An example of a Killing horizon is then given by 

\begin{equation} X^2 = V^2 \ , \end{equation}

\noindent and this null surface shares a one parameter family of generators 
with the black hole horizon. The Killing horizon is also the boundary of the 
causal past of a timelike curve ending up on the singular circle at $(t,r,\phi ) 
= (\pi/2, 1, 0)$, hence it is a non-compact observer dependent event horizon. 
The black hole event horizon is the envelope of the set of all such observer 
dependent event horizons \cite{Mann}. In the interior of the black hole we 
expect to find closed trapped surfaces, surrounded by an apparent horizon lying 
well inside the event horizon. This is indeed so \cite{holstpeldan}.

In the following we will rely on two coordinate description of the toroidal black 
hole. A coordinate system covering the region $X^2 + Y^2 > V^2$, with $U > 0$, in 
anti-de Sitter space is   

\begin{equation} \begin{array}{lll} 
X = r\cosh{t}\cos{\phi}  & \ & 0 < r < \infty \\
Y = r\cosh{t}\sin{\phi} & \ & 0 \leq \phi < 2\pi \\
Z = \sqrt{r^2+1}\sinh{\gamma} & \ & - \infty < t < \infty \\
U = \sqrt{r^2+1}\cosh{\gamma} & \ & \\
V = r\sinh{t} & \ & \hspace{5mm} . \end{array}
\end{equation}

\noindent The metric in these coordinates is 

\begin{equation} ds^2 = r^2(-dt^2 + \cosh^2{t}d\phi^2) + \frac{dr^2}{r^2+1} 
+ (r^2+1)d\gamma^2 \ . \label{ext} \end{equation}

\noindent The manifest Killing vectors are 

\begin{equation} J_{ZU} = \partial_\gamma \hspace{10mm} J_{XY} = 
\partial_\phi \ . \end{equation}

\noindent The black hole is obtained by making $\gamma$ periodic, and the 
coordinate system then covers its exterior.  

A different coordinate system, covering the region $U^2 > Z^2$, 
$U > 0$, is 
 
\begin{equation} \begin{array}{lll} 
X = \sqrt{r^2-1} \cos{\chi} & \ & 1 < r < \infty \\
Y = \sqrt{r^2-1}\sin{\chi} & \ & 0 \leq \chi < 2\pi \\
Z = r\sin{\tau}\sinh{\phi} & \ & 0 < \tau < \pi \\
U = r\sin{\tau}\cosh{\phi} & \ & \\
V = r\cos{\tau} & \ & \hspace{5mm} . \end{array}
\end{equation}

\noindent These coordinates cover an entire region with spacelike $J_{ZU}$. 
The metric is 

\begin{equation} ds^2 = r^2(-d\tau^2 + \sin^2{\tau}d\phi^2) + 
\frac{dr^2}{r^2-1} + (r^2-1)d\chi^2 \ . \label{global} \end{equation} 

\noindent We have two manifest Killing vectors, namely 

\begin{equation} J_{ZU} = \partial_\phi \hspace{10mm} J_{XY} = \partial_\chi \ . 
\label{Killing1} \end{equation}

\noindent The black hole is obtained by making $\phi$ periodic. The Misner 
singularities are then at $\sin{\tau} = 0$, and the event 
horizon is at $r = 1/\sin{\tau}$. 

Further local Killing vectors include 

\begin{equation} J_{ZV} = - \sinh{\phi}\partial_\tau 
+ \cosh{\phi}\cot{\tau}\partial_\phi \hspace{4mm} J_{UV} = - \cosh{\phi}\partial_{\tau} + 
\sinh{\phi}\cot{\tau} \partial_\phi \ . \label{Killing2} \end{equation}

\noindent Together with $J_{ZU}$ they form a local $SO(2,1)$ algebra. Since they 
do not commute with $J_{ZU}$ they are not globally defined in the black hole 
spacetime, but we record them here since they will be of interest in section 
5. 

\vspace{1cm}

{\bf 3. Analytic relatives of Schwarzschild}

\vspace{5mm}

\noindent A useful trick, capable of producing new solutions from old, 
is to complexify a solution of Einstein's equations, and then take a new 
real slice of the result \cite{Brill}. Perhaps it is more than a trick, 
but if so this does not concern us now. 

What does concern us is that the 
Schwarzschild solution has three natural relatives, obtainable by 
analytic continuation in this way. This gives us four solutions altogether, 
given in terms of one of the two functions 

\begin{equation} V_{\pm }(r) = \pm 1 - \frac{2m}{r} - \frac{\lambda r^2}{3} 
\ . \end{equation}

\noindent For the moment the cosmological constant $\lambda$ is kept arbitrary. 
As long as only the local geometry matters the solutions are 

\begin{eqnarray} ds^2 = - V_+(r)dt^2 + \frac{dr^2}{V_+(r)} + r^2(d\theta^2 
+ \sin^2{\theta}d\phi^2) =  \nonumber \\ 
\ \label{Sch} \\ 
\hspace{2cm} = - V_+(r)dt^2 + \frac{dr^2}{V_+(r)} + 
r^2(\mbox{sphere}) \nonumber \end{eqnarray}  
 
\begin{eqnarray} ds^2 = - V_-(r)dt^2 + \frac{dr^2}{V_-(r)} + r^2(d\theta^2 
+ \sinh^2{\theta}d\phi^2) =  \nonumber \\ 
\ \label{top} \\ 
\hspace{2cm} = - V_-(r)dt^2 + \frac{dr^2}{V_-(r)} + 
r^2(\mbox{hyperbolic plane}) \nonumber \end{eqnarray}  
 
\begin{eqnarray} ds^2 = r^2(- dt^2 + \cosh^2{t}d\phi^2) + \frac{dr^2}{V_+(r)} 
+ V_+(r)d\gamma^2 =  \nonumber \\ 
\ \label{bubbla} \\ 
\hspace{2cm} = r^2(\mbox{de Sitter}) + \frac{dr^2}{V_+(r)} 
+ V_+(r)d\gamma^2 \nonumber \end{eqnarray}

\begin{eqnarray} ds^2 = r^2(- d\tau^2 + \sin^2{\tau}d\phi^2) + 
\frac{dr^2}{V_-(r)} + V_-(r)d\chi^2 =  \nonumber \\ 
\ \label{bh} \\ 
\hspace{18mm} = r^2(\mbox{anti-de Sitter}) + \frac{dr^2}{V_-(r)} 
+ V_-(r)d\chi^2 \ . \nonumber \end{eqnarray}
  
\noindent To go from the metric (\ref{Sch}) to (\ref{bh}), say, we use 

\begin{equation} t \rightarrow \chi \ , \ \ r \rightarrow ir \ , \ \ 
\theta \rightarrow \tau \ , \ \ \phi \rightarrow i\phi \ , 
\ \ m \rightarrow -im \ . \end{equation}

\noindent The others are similarly related.  
The local isometry group of these metrics is ${\bf R}\times SO(3)$ for 
the first case, and ${\bf R}\times SO(1,2)$ for the other three. Since 
the coordinate ranges are as yet undecided, the global isometry groups 
and the topologies of the underlying spacetimes remain to be found. 

The metric (\ref{Sch}) is the Schwarzschild, or Schwarzschild-(anti)-de 
Sitter, solution. The metric (\ref{top}) 
is not interesting as it stands, but if the isometries are used to turn 
the hyperbolic plane into a closed Riemann surface it describes the 
topological black holes \cite{4d, BLP}. The metric (\ref{bubbla}) describes  
a completely regular spacetime, once the coordinate singularity at $V_+(r) = 0$ 
has been dealt with appropriately \cite{Ehlers}. It is referred to as a 
``bubble of nothing'', and will be discussed in the next section. The metric 
(\ref{bh}) has been called an ``anti-bubble'' \cite{Astefan}, and is the topic 
of our section 5.  

\vspace{1cm}

{\bf 4. Bubbles of nothing}

\vspace{5mm}

\noindent We will now discuss the ``bubble of nothing'' metric 
(\ref{bubbla}) in some detail, although we will keep the discussion fairly 
brief because many of the things we will say can be found in the literature 
already, also for the case $\lambda < 0$ \cite{Birmingham, Ross}. When 
$m = 0$ it reduces to the anti-de Sitter metric (\ref{ext}), which describes 
the exterior of a toroidal black hole if $\gamma$ is periodic \cite{Cai}. 
For all other values of $m$ there is a coordinate singularity at $V(r) = 0$, 
and a carefully adjusted periodicity in $\gamma$ is necessary if one is to 
obtain an everywhere regular spacetime \cite{Ehlers}. It has topology 
${\bf R}^3\times {\bf S}^1$. 

To be precise, introduce a new radial coordinate $\sigma$ through 

\begin{equation} \sigma^2 \equiv V_+(r) = 1 - \frac{2m}{r} - 
\frac{\lambda r^2}{3} 
= (r-r_+)\frac{r^2 + r_+r + r_+^2 + 1}{r} \ . \end{equation}

\noindent In the last step we actually assumed that $\lambda < 0$, and 
defined the largest positive root $r_+$. Then 

\begin{equation} dl^2 = \frac{dr^2}{V_+(r)} + V_+(r)d\gamma^2 = 
\frac{4d\sigma^2}{V_+^{\prime \ 2}} + \sigma^2d\gamma^2 \ . \end{equation}

\noindent This metric will be regular at the origin if and only if 
$\gamma$ is periodic with 

\begin{equation} 0 \leq \gamma < \frac{4\pi r_+}{3r_+^2 + 1} \ . 
\end{equation}   

\noindent The periodicity in $\phi$ is arbitrary, although a period 
which is a multiple of $2\pi$ is preferred because it allows a globally 
defined isometry group $SO(2)\times SO(2,1)$.  

Consider the case $\lambda = 0$. The first detailed discussion 
of this spacetime was given by Witten \cite{Witten}, who was interested in 
its five dimensional version as an example of an asymptotically flat solution 
with vanishing mass. It is not asymptotically Minkowski however; as 
noted by Aharony et al. \cite{Aharony} it does not admit a sensible \scri . 
One can understand this either as a consequence of a general theorem---an 
asymptotically simple spacetime cannot have a null infinity with topology 
${\bf T}^2\times {\bf R}$ \cite{Claudel}---or by inspection of the special 
case $m = 0$. When $m = 0$ the spacetime is obtained by identifying points 
in Minkowski space, using a translation. For Minkowski space \scri$^+$ has 
topology ${\bf S}^2 \times {\bf R}$. Spatial translations act along the null 
generators on \scri$^+$, except for an equator's worth of generators that are 
left untouched. When the identifications are carried through this will 
result in closed null curves on \scri .

When $\lambda < 0$ the situation is very different. In this case the intrinsic 
geometry of \scri \ is independent of $m$. In the coordinates used to express 
the metric, we choose the conformal factor 

\begin{equation} \Omega = \frac{1}{r} \ . \end{equation}

\noindent This ensures that \scri \ is a totally geodesic surface at 
$\Omega = 0$, with respect to the unphysical metric $\hat{g}_{ab} = \Omega^2
g_{ab}$. Its intrinsic metric is 

\begin{equation} d\hat{s}^2_{\Omega = 0} = - dt^2 + \cosh^2{t}d\phi^2 
+ d\gamma^2 \ . \label{mscri} \end{equation}

\noindent Although this is de Sitter space times a circle, the 
conformal properties are quite different from that of de Sitter space 
itself \cite{I}.  

\begin{figure}
        \centerline{ \hbox{
               \epsfig{figure=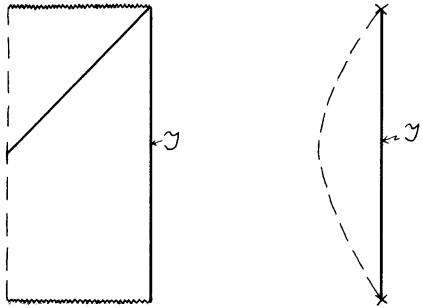,width=6cm}}}
        \caption{\small Penrose diagram of the adS black hole (left), and of 
the corresponding B metric. Each point is a torus, except for the points on 
the dashed lines which are circles. The observer dependent horizons cannot 
be seen here.}
        \label{fig:3}
\end{figure} 

The Penrose diagrams of these spacetimes are given 
in fig.\ \ref{fig:3}. They do not make full justice to the causal structure, 
because an observer may end up anywhere on the singular circle that terminates 
\scri . This gives rise to observer dependent non-compact Killing event 
horizons \cite{Mann, Aharony}, which can be seen only if we rotate the Penrose 
diagram around a vertical axis. This latter operation also explains why 
the $m \neq 0$ solutions are referred to as ``bubbles of nothing'' in the 
literature: the vertical axis of rotation should be placed so that the 
dashed curve turns into a de Sitter hyperboloid, and there is simply nothing 
in the center of the picture \cite{Aharony}.  

What is the mass? The proper definition of conserved charges for asymptotically 
anti-de Sitter spacetimes has been the subject of a large body of recent research, as 
one can see by consulting, say, Gibbons et al. \cite{Gibbons} or Papadimitriou 
and Skenderis \cite{Kostas}. Our case is a relatively simple one since its 
conformal boundary is even dimensional and conformally flat. 
Therefore we will rely on the definitions by Ashtekar 
and Magnon \cite{Ashtekar}, which are based on \scri . When its cross 
sections are closed 2-manifolds, \scri \ will always admit conserved 
charges. The explicit expressions were given by Ashtekar and Magnon---who 
assumed that the topology is ${\bf S}^2\times {\bf R}$, but their charges 
remain conserved in our case. They are expressed in terms of an asymptotic 
Killing vector together with the rescaled Weyl tensor 

\begin{equation} K_{abcd} = \frac{1}{\Omega}\hat{C}_{abcd} = \Omega 
C_{abcd} \ . \end{equation}   

\noindent More precisely we need its electric and magnetic parts

\begin{equation} E_{ab} = K_{acbd}n^cn^d \hspace{8mm} B_{ab} = 
\star K_{acbd}n^cn^d \ , \hspace{8mm} n_a = \nabla_a\Omega \ . 
\end{equation}

\noindent For the metric at hand we find the non-vanishing components  

\begin{equation} E_{tt} = m \hspace{8mm} E_{\phi \phi} = - m\cosh^2{t} 
\hspace{8mm} E_{\gamma \gamma} = 2m \ . \end{equation}

\noindent The magnetic part vanishes identically, which means that \scri \ 
is conformally flat.

The conserved charges are defined by 

\begin{equation} Q = \oint E_{ab}\xi^adS^b \ , \end{equation}

\noindent where the integral is taken over a toroidal cross section of 
\scri , and $\xi$ is an asymptotic Killing vector, hence a conformal 
Killing vector of the metric (\ref{mscri}). But the most general such 
vector field available is a linear combination of the four vector fields 

\begin{eqnarray} \xi_1 = \cos{\phi}\partial_t - \tanh{t}\sin{\phi}\partial_\phi 
\hspace{10mm} \xi_3 = \partial_\phi \ \ \nonumber \\
\ \\
\xi_2 = \sin{\phi}\partial_t + \tanh{t}\cos{\phi}\partial_\phi \hspace{10mm} 
\xi_4 = \partial_\gamma \ . \nonumber  \end{eqnarray}

\noindent Provided that $\phi$ is periodic with period $2\pi n$ all of these 
exist globally, but it is easily checked that all the conserved charges vanish. 
Hence the mass of these bubbles of nothing is zero, even though superenergies 
and quasi-local masses are non-zero. The mass of the toroidal black hole is 
also zero. This reminds us of the C-metric, which has zero ADM mass 
even though its Bondi news tensor is non-vanishing \cite{Cmetric}.

\vspace{5mm}

{\bf 5. Black hole spacetimes}

\vspace{5mm}

\noindent Now consider the metric (\ref{bh}), with $\lambda = - 3$: 

\begin{equation} ds^2 = r^2(- d\tau^2 + \sin^2{\tau}d\phi^2) + 
\frac{dr^2}{V_-(r)} + V_-(r)d\chi^2 \ , \hspace{5mm} V_-(r) = 
r^2 - 1 - \frac{2m}{r} \ . \end{equation}

\noindent The coordinates $\tau$ and $\phi$ cover a maximal causal diamond in 
a 1+1 dimensional anti-de Sitter space. When $m = 0$ and the coordinate $\phi$ 
is made periodic the metric is that of the quotient black hole, in the 
coordinate system where its 
metric takes the form (\ref{global}). The coordinate $\chi$ must be 
periodic in order to avoid a coordinate singularity at $V_-(r) = 0$. 
The appropriate period is 

\begin{equation} 0 \leq \chi < \frac{4\pi r_+}{3r_+^2 - 1} \ , \end{equation}

\noindent where $r_+$ is the largest real root of $V_-(r)$. The range of 
the radial coordinate is then $r > r_+$. It is not hard to see that 

\begin{equation} 2m = r^3_+ - r_+ \ . \end{equation}

\noindent Hence $m$ can have either sign, but we make the 
restriction that $r_+$ be positive because if $r$ is allowed to take the 
value zero a curvature singularity arises. Since $r_+$ is also the largest 
root we find that 

\begin{equation} m \geq - \frac{1}{3\sqrt{3}} \hspace{12mm}
 r_+ \geq \frac{1}{\sqrt{3}} \ . \end{equation}

\noindent The locally anti-de Sitter case is at $r_+ = 1$. 
With this restriction the curvature remains bounded everywhere, but since 
we make $\phi$ periodic there 
are Misner singularities at $\tau = 0$ and $\tau = \pi$. 

The Misner singularities will be present also on the conformal boundary 
\scri . Indeed, as a three dimensional manifold \scri \ is independent of 
$m$, and the discussion of the anti-de Sitter case $m = 0$ can be taken 
over {\it verbatim} \cite{I}. In particular there will be an event horizon 
bounding the region of spacetime that can be seen from \scri . It can be 
located by solving for radial null geodesics ending up at the Misner 
singularity at $\tau = \pi$. This means that a point with coordinates 
$(r, \tau)$ belongs to the horizon only if 

\begin{equation} \int_\tau^\pi d\tau = \int_r^\infty \frac{dr}{r\sqrt{V_-(r)}} 
= \int_r^\infty \frac{dr}{\sqrt{r}\sqrt{r^3 - r - 2m}} \ . \end{equation}

\noindent This is an elliptic integral of the first kind, degenerating to 
an elementary integral at $r_+ = 1/\sqrt{3}$, $r_+ = 1$, and 
$r_+ = 2/\sqrt{3}$. 

\begin{figure}[ht]
        \centerline{ \hbox{
                \epsfig{figure=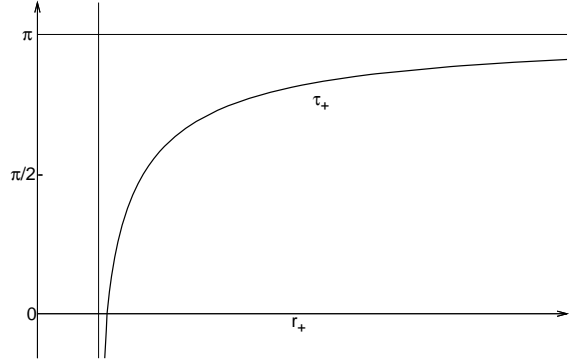,width=55mm,angle=270}}}
        \caption{\small This is 
$\tau_+$ as a function of $r_+$; these quantities are defined 
in the text.}
        \label{fig:fig-mass}
\end{figure}

We are especially interested in $\tau_+$, which is the value of $\tau$ 
at which the event horizon is born. Clearly 

\begin{equation} \tau_+ = \pi - \int_{r_+}^\infty \frac{dr}
{\sqrt{r}\sqrt{r^3 - r - 2m}} \ . \end{equation}

\noindent The numerical solution is displayed in Fig. \ref{fig:fig-mass}. 
The fact that 

\begin{equation} \tau_+ \sim \pi - \frac{\mbox{constant}}{r_+} 
\end{equation}

\noindent in the limit of large $r_+$ is easily verified analytically. 
Note also that $\tau_+$ diverges to negative infinity at $r_+ = 1/\sqrt{3}$. 
Negative values of $\tau_+$ are not realized; when they occur they 
mean that no complete spatial slice, going all the way down to $r_+$, is 
visible from \scri .    
 
\begin{figure}[ht]
        \centerline{ \hbox{
                \epsfig{figure=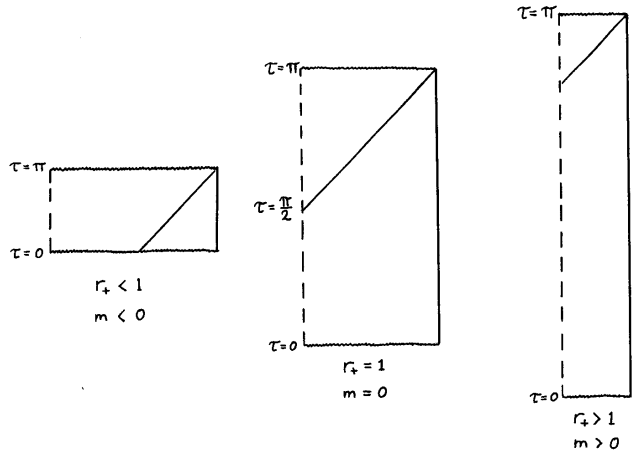,width=9cm}}}
        \caption{\small Penrose diagrams for the black hole family. They depend on 
$r_+$, and hence on $m$, in a characteristic way.}
        \label{fig:5}
\end{figure}

The causal structure of these spacetimes can be displayed by Penrose diagrams 
where each point is a torus. See Fig. \ref{fig:5}. As in the previous cases these 
diagrams do less than full justice to the causal structure because 
the observer dependent horizons cannot be seen. On the other hand the 
dependence of the causal structure on $r_+$, and implicitly on $m$, 
is seen clearly. When $m$ is large the diagram is very tall, and when 
$m$ is close to its lower bound it becomes a rectangle whose width is 
much larger than its height. 

Locally the isometry group of the solution is ${\bf R}\times SO(2,1)$, 
just as for the bubbles of nothing. However, on examination one finds 
that the local Killing vectors are given in terms of the intrinsic 
coordinates by eqs. (\ref{Killing1}-\ref{Killing2}). But the global 
structure of the solution breaks the symmetries described by eqs. 
(\ref{Killing2}), so the global isometry group is only $SO(2)\times 
SO(2)$. The case $m = 0$ is exceptional in that it has more symmetries, 
not given by these Killing vectors. 

The Ashtekar-Magnon conserved charges are therefore very easy to compute. 
Just as for the bubble of nothing, the electric part of the rescaled Weyl 
tensor is diagonal (and its magnetic part vanishes), but this time there 
are no global Killing vector having a ``time'' component, and so the 
Ashtekar-Magnon conserved charges are zero. None of them 
seem to deserve the name ``mass''.   

\vspace{1cm}

{\bf 6. Summary}

\vspace{5mm} 

\noindent We have observed that the toroidal anti-de Sitter black hole 
\cite{4d} can be regarded as a member of two different one parameter 
families of spacetimes, both of them obtainable by analytic continuation 
from the Schwarzschild-anti-de Sitter 
solution, if suitable topological identifications are made. One of 
these families contains regular spacetimes called ``bubbles of nothing'', 
and they have been discussed in some detail elsewhere \cite{Ehlers, Witten, 
Aharony}.  The other family consists of black hole spacetimes, with 
Misner singularities to terminate \scri , but without curvature singularities. 
The global isometry group of the latter is just $SO(2)\times SO(2)$, and 
the causal structure is similar to that of the anti-de Sitter black hole 
itself.  Both families have vanishing conserved charges in the sense of 
Ashtekar and Magnon \cite{Ashtekar}, hence the mass (if any) is zero. We 
argued that this is unsurprising. 

We think that these results are of some interest as examples of model 
spacetimes; perhaps they can find some use in the context of the adS/CFT 
correspondence. However, although the locally anti-de Sitter black hole 
exists in all dimensions, it is not obvious to us how to generalize the 
entire black hole family to dimensions higher than four.

\end{document}